\documentclass[twocolumn,english,superscriptaddress,
preprintnumbers,amsmath,amssymb,floatfix]{revtex4}

\usepackage[T1]{fontenc}
\usepackage[latin9]{inputenc}
\usepackage{dcolumn}
\usepackage{bm}
\usepackage{graphicx}
\usepackage{color}
\usepackage{esint}
\usepackage{babel}
\usepackage{amsfonts}
\usepackage{slashed}
\usepackage{bm}
\usepackage{amsmath}
\usepackage{epsfig}
\usepackage{enumerate}
\usepackage{hyperref}
\usepackage{doi}
\def\be{\begin{equation}}
\def\ee{\end{equation}}

\begin{document}

\title{Proximity induced time-reversal topological superconductivity in Bi$_2$Se$_3$ films without phase tuning}

\author{Oscar E. Casas}
\affiliation{Departamento de F\'{\i}sica, Universidad Nacional de Colombia, Bogot\'a,
 Colombia}
\affiliation{Departamento de F{\'i}sica Te{\'o}rica de la Materia Condensada C-V, Condensed Matter Physics Center (IFIMAC) and Instituto Nicol\'as  Cabrera,  Universidad Aut{\'o}noma de Madrid, E-28049 Madrid, Spain} 

\author{Liliana Arrachea}
\affiliation{International Center for Advanced Studies, Escuela de Ciencia y Tecnolog\'{\i}a, Universidad Nacional de San Mart\'{\i}n-UNSAM, Av 25 de Mayo y Francia, 1650 Buenos Aires, Argentina}

\author{William J. Herrera}
\affiliation{Departamento de F\'{\i}sica, Universidad Nacional de Colombia, Bogot\'a,
 Colombia}

\author{Alfredo Levy Yeyati}
\affiliation{Departamento de F{\'i}sica Te{\'o}rica de la Materia Condensada C-V, Condensed Matter Physics Center (IFIMAC) and Instituto Nicol\'as  Cabrera,  Universidad Aut{\'o}noma de Madrid, E-28049 Madrid, Spain} 

\date{\today}

\begin{abstract}
Many proposals to generate a time-reversal invariant topological superconducting phase are based on imposing a $\pi$ phase difference between the superconducting
leads proximitizing a nanostructure. We show that this phase can be induced on a thin film of a topological insulator like Bi$_2$Se$_3$ in proximity to
a {\it single} s-wave superconductor. In our analysis we take into account the parity degree of freedom of the electronic states which is not included in effective Dirac-like surface theories. We find that the topological phase can be reached when the induced interparity pairing dominates over the intraparity one. Application of an electric field perpendicular to the film extends the range of parameters where the topological phase occurs. 
\end{abstract}


\maketitle

\section{Introduction} 
The interest in topological phases of matter and, in particular, in topological superconductors (TSs) has not ceased to grow \cite{review}.
In addition to their fundamental interest, TSs are predicted to host topologically protected Majorana zero modes (MZM) at the edges
 with potential applications in future quantum technologies \cite{QT}. 

\begin{figure}[t]
\includegraphics[width=\columnwidth]{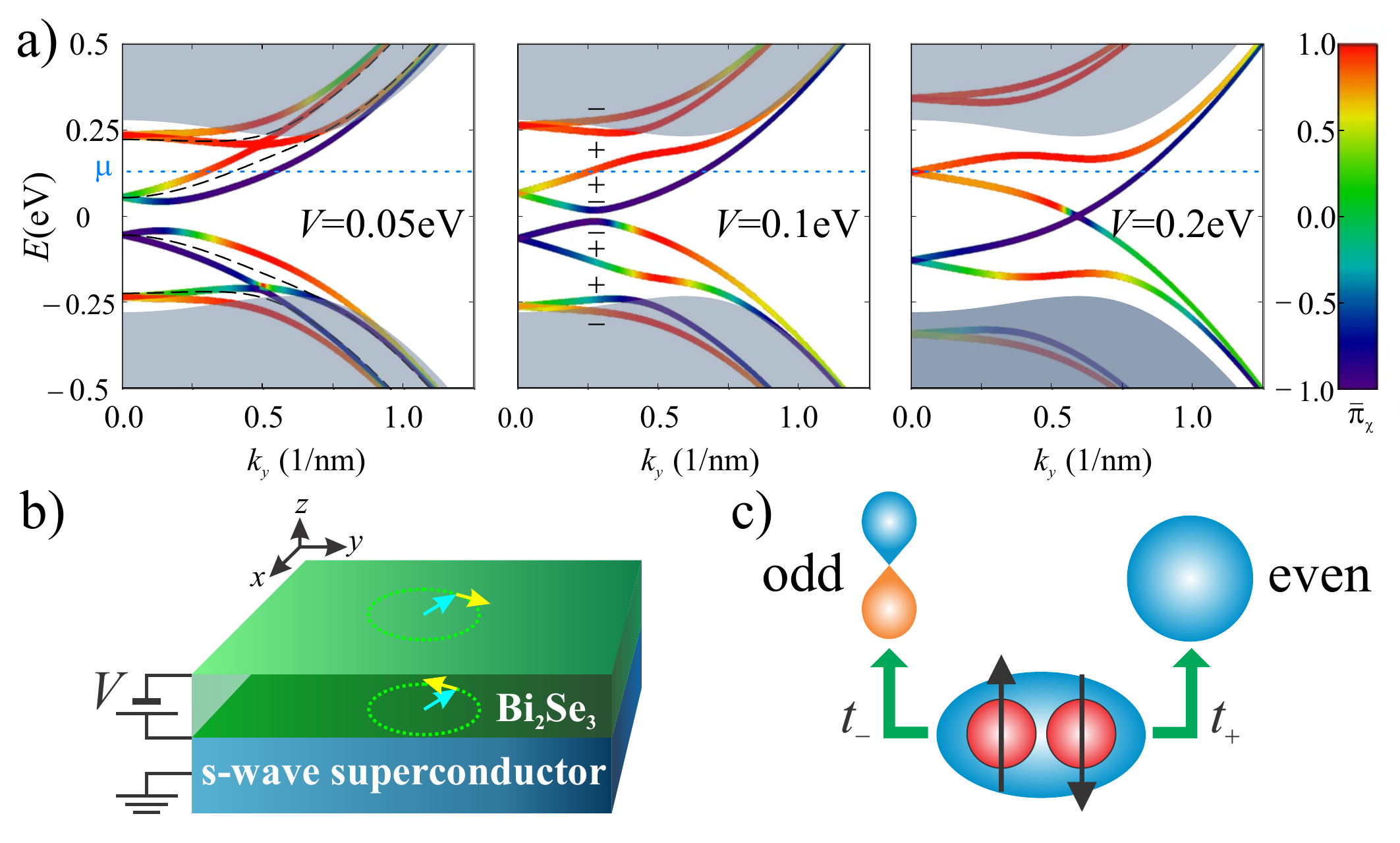}
\caption{a) Surface states bands in a thin Bi$_2$Se$_3$ film in the presence of an electric field, controlled by the biasing potential $V$ between the top and bottom surfaces. The bands are helicity degenerate for $V=0$ (dashed lines in left panel) but the degeneracy is broken for finite $V$. The signs in the middle panel indicate the bands helicity and the color scale of the lines is set by the normalized relative weight, $\bar{\pi}_{\chi}$, of the surface states on the two parity sectors (with
$\bar{\pi}_{\chi}=2\pi_{\chi}/(1+ \pi_{\chi}^2)$, where $\pi_{\chi}$  is the relative weight defined in the main text). 
The gray areas indicate the regions 
for the bulk states and the dashed horizontal line indicates the position of the chemical potential.
b) Geometry considered for analyzing the 
proximity effect. c) Schematic representation of the interparity pairing which can be induced from the $s$-wave superconductor.}
\label{figure1}
\end{figure}

Although topological superconductivity is expected to occur spontaneously in certain compounds like Sr$_2$RuO$_4$ \cite{sr2ruo4}; actual vigorous experimental progress is coming from the side of artificial nanostructures. In particular,
clear signatures of MZMs have been demonstrated in hybrid nanostructures combining semiconducting nanowires with strong spin orbit 
(like InAs or InSb) and conventional superconductors \cite{exps-nanowires-1,exps-nanowires-2,exps-nanowires-3,exps-nanowires-4,exps-nanowires-5}.

As in the case of other proposals based on arrays of magnetic impurities \cite{impurities} these platforms constitute a realization of 
broken-time reversal (symmetry class D) 1D topological superconductivity.  Although  
the time-reversal counterpart or class DIII superconductivity has attracted great theoretical interest \cite{oreg-review},
its actual realization is still an experimental challenge. The zero energy excitations in this class of TSs are  Kramers pairs of Majorana modes. 
While their braiding properties appear to be path dependent 
\cite{stern,stern1}, they exhibit other exotic transport \cite{tritops-ber, jose1} and spin \cite{kesel, cam,B-schaffer-2,fractio} properties which render them objects of fundamental interest.

Intrinsic DIII superconductivity in two and three dimensional systems has been discussed in the literature (see for instance Refs. \cite{zhang,fu-berg,Schmalian}) but also in this case most theoretical proposals have been focused on proximitized nanostructures.  These, in general, require two basic ingredients: a multichannel or multiband electronic structure
and a mechanism for inducing opposite pairing amplitudes on these channels \cite{no-go-theorem}. 
These include Rashba nanowires proximitized by a d-wave \cite{Wong} or an Iron-based superconductor with
$s_{\pm}$ pairing symmetry \cite{Zhang-Kane-Mele}; or two parallel nanowires with interwire pairing \cite{Flensberg,Klinovaja,Ebisu} or subject to
opposite Zeeman fields \cite{Reeg}. Another scenario is spin orbit and many body interactions in proximity with ordinary superconductivity \cite{haim,yuval1}.
 Induction of the DIII phase on the edge or surface states of a 2D or a 3D 
topological insulator (TI) has also been considered \cite{Fu-Kane,Santos,
Klinovaja,Liu,Schrade,Black-Schaffer}. Refs. \cite{Liu,Black-Schaffer} suggest that for the case of a thin 3D TI film reaching the DIII phase requires forming a S/TI/S junction and imposing a $\pi$-phase difference. These studies are based on effective 2D models describing the surface states on the 3D TI film.
 
In the present work we propose a new approach for the case of proximitized 3DTI thin films. In contrast to previous works which start from the projected 2D theory, we use a
3D model which keeps track of the parity degree of freedom. We show that the 
DIII-TS phase may arise naturally by proximity to a {\it single} s-wave
superconductor when considering the presence of interparity pairing. We further show that the inclusion of an external electric field, breaking inversion symmetry, helps to stabilize this topological phase.

The main ingredients of the proposed mechanism are illustrated in Fig. \ref{figure1}. The surface states of a 3D TI are characterized by a well defined helicity, i.e. they are eigenstates of the helicity operator $\hat{h} = (\bm{\sigma}\times {\bf k}_{\parallel})_z/k_{\parallel}$, where ${\bf k}_{\parallel}$ is the wave vector parallel to the surface and $\sigma_{\alpha}$ are Pauli matrices in spin space \cite{kp-Zhang}.  
In addition, these states are also characterized by a certain parity pseudospin, which depends on the surface orientation and on the material. For instance, in films of the Bi$_2$Se$_3$ family grown along the $c$ axis,
states on opposite surfaces have opposite helicities and opposite parity pseudospin
\cite{Liu-Zhang}. 
In sufficiently thin films, the surface states corresponding to oposite sides are not fully decoupled but hybridize to some extent \cite{Bi2Se3-films,Niu} and the helicity degeneracy can be broken by an electric field perpendicular to the film, as illustrated in Fig. \ref{figure1}(a).
Therefore, when one of the surfaces is in contact with a superconductor as in Fig. \ref{figure1}(b), superconductivity is induced into the two surfaces 
in both parity channels, as well as interparity, as schematically depicted in Fig. \ref{figure1}(c). Interestingly, the interparity component induced on each helical channel tipically have opposite signs. 
Our goal is to show that for the case of an s-wave superconductor, the TS phase can be reached provided that the interparity component is large enough and
inversion symmetry is broken. 

\section{Model for a TI film and proximity effect} 

The low energy and long wavelength electronic properties of a TI of the
Bi$_2$Se$_3$ family can be described by 
the $\bf{k}\bf{\cdot}\bf{p}$ Hamiltonian introduced in Ref. 
\cite{kp-Zhang} in a basis of four states which are combinations of $p_z$ 
orbitals on the Bi and Se sites with even and odd parities. For analyzing the 
proximity effect it is convenient to perform
a unitary transformation \cite{Liu-Zhang,Silvestrov} with respect to the model
in Ref. \cite{kp-Zhang} [see discussion in the Supplementary Material (SM), Ref. \onlinecite{supplementary}], which yields
\begin{eqnarray}
H^{3D} &=& {\cal M}({\bf k}) \tau_z \otimes \sigma_0 + A_1 k_z \left(\tau_y \otimes \sigma_0\right) \nonumber \\
&&  - A_2 \left[k_x \left(\tau_x \otimes \sigma_y\right) - k_y \left(\tau_x \otimes \sigma_x\right) \right],\;
\end{eqnarray} 
where ${\cal M}({\bf k}) = M_0 - B_1 k_z^2 - B_2 (k_x^2+k_y^2)$ \cite{parameters}, while $\tau_{\alpha}$
are Pauli matrices operating in parity space. 
This Hamiltonian commutes with the helicity
operator, leading to the properties of the surface states commented above. 

In order to describe the proximitized thin film we now switch into a tight-binding (TB) description of the electronic structure. For this purpose we follow
Ref. \cite{acero} and introduce a cubic lattice with parameter $a \sim 1$ nm oriented along the $c$-axis and consider the ${\bf k}{\bf \cdot}{\bf p}$ Hamiltonian as a long-wavelength expansion of this TB model. We shall consider the case of 
films of thickness $L_z = N_z a$ and impose periodic boundary conditions on the $x,y$ directions. In the basis $\psi_{k_{\parallel},i} =
(c_{k_{\parallel},i+\uparrow}, c_{k_{\parallel},i+\downarrow}, c_{k_{\parallel},i-\uparrow},
c_{k_{\parallel},i-\downarrow})^T$, where $c^{\dagger}_{k_{\parallel},i\tau\sigma}$
creates an electron with parallel momentum ${\bf k}_{\parallel}$ on the $i$-layer within
the film, parity $\tau$ and spin $\sigma$. The TB model adopts the form
$\hat{H}^{TB} = \sum_{k_{\parallel},ij} \psi^{\dagger}_{k_{\parallel},i} \hat{\cal H}({\bf k}_{\parallel})_{ij} \psi_{k_\parallel,j}$ where
\begin{eqnarray}
&&\hat{\cal H}({\bf k}_{\parallel})_{ij} = \epsilon({\bf k}_{\parallel}) \left(\tau_z \otimes \sigma_0 \right) \delta_{ij} + \frac{A_2}{a} \left[\sin k_y a \left(\tau_x \otimes \sigma_x\right) \right. \nonumber\\
&&-\left. \sin k_x a \left(\tau_x \otimes \sigma_y\right) \right] \delta_{ij}+
\frac{B_1}{a^2} \left(\tau_z\otimes\sigma_0\right) (\delta_{ij-1}+\delta_{ij+1}) \nonumber\\
&&-\frac{iA_1}{2a} \left(\tau_y\otimes\sigma_0\right) (\delta_{ij-1}-\delta_{ij+1}),
\end{eqnarray}
with $\epsilon({\bf k}_{\parallel})= M_0 - 2\left[B_2(2 - \cos k_x a - \cos k_y a)+B_1\right]/a^2$. 
Within this model the eigenstates are again helicity degenerate (with the helicity operator properly extended to the discrete case) but this degeneracy is broken when an electric field along the $z$ direction, $\hat{\cal V}_{ij} = 2V(i-(N_z+1)/2)/(N_z-1) \delta_{ij}$, is introduced.

To include the effect of induced pairing correlations on the film we consider the Bogoliubov de Gennes (BdG) Hamiltonian,
expressed in the basis $\Psi^{\dagger }_j(\mathbf{k}_{\parallel })=\left( \psi_{k_{\parallel},j},-i\sigma _{y}\psi_{-k_{\parallel},j}\right) $. It reads
\begin{equation}
\hat{\cal H}^{BdG}(k_{\parallel})_{ij}= \left(\hat{\cal H}({\bf k}_{\parallel})_{ij}+\hat{\cal V}_{ij} -\mu \right)\otimes \eta_z + \hat{\Delta}_{ij} \otimes \eta_x,
\end{equation}
where $\mu$ is the chemical potential, and $\eta_j$ are Pauli matrices in the particle-hole space.
Although this model allows for more general configurations we shall focus in this work in the case 
depicted in Fig. \ref{figure1}(b), where pairing is induced on the
$i=1$ layer only, i.e. $\hat{\Delta}_{ij} = \hat{\Delta}_1 \delta_{i1} \delta_{ij}$. $ \hat{\Delta}_1$ has 
 {\it intra} ($\Delta_{\pm}$) and {\it inter} ($\Lambda$) parity components, 
\begin{equation}
\hat{\Delta}_1 =  \frac{ \Delta_{+} }{2} \left(\tau_0+ \tau_z \right)+\frac{ \Delta_{-} }{2} \left(\tau_0- \tau_z \right)+\Lambda \tau_x .
\label{pairing}
\end{equation}
The pairing potentials depend on the coupling of the TI with the
superconductor underneath. As discussed in the SM \cite{supplementary}, they would typically have the form
$\Delta_{\pm}= \pi \rho_F t_{\pm}^2 $ and $\Lambda= \pi t_+ t_- \rho_F$, where $\rho_F$ is the superconductor Fermi level density of states and $t_{\pm}$
are hopping parameters coupling the TI orbitals with parity $\pm$ and the superconductor. These parameters may have oposite signs. In particular, when the ordinary superconductor is contacted to the bottom of the film one expects $t_+ t_- <0$, which implies that $\Lambda$ has an overall sign with respect to $\Delta_{\pm}$.
On the contrary, when the superconductor is contacted to the top surface, $t_+t_->0$ and thus $\Delta_{\pm}$ and $\Lambda$ have the same sign. 

It should be stressed that the above expressions are fully compatible with time-reversal symmetry.
Regarding the size of $\Lambda$, while a non-interacting model suggests $\Lambda \sim \sqrt{\Delta_+\Delta_-}$, the
presence of moderate local Coulomb repulsion on the Bi and Se sites would yield
the condition  $\Lambda > \sqrt{\Delta_+\Delta_-}$ which is necessary for 
stabilizing the DIII-TS phase as we show below.

\section{Topological invariant} 

In the limit of weak coupling, 
the topological character of the proximitized TI film can be fully determined
by the normal electronic properties at the Fermi level \cite{Zhang-invariant}. 
The $Z_2$ topological invariant introduced in Ref. \cite{Zhang-invariant} is given by
\begin{equation}
N = \prod_n  \left(\mbox{sign} \langle \psi_n(k_{F,n})|\mathcal{T }\hat{\Delta}^{\dagger}|\psi_n(k_{F,n})\rangle\right)^{m_n} \;,
\label{invariant}
\end{equation}
where $\mathcal{T}= \tau_0 \otimes i \sigma_y K$ with $K$ denoting complex conjugation, is the time-reversal operator, $n$ runs over all bands crossing the Fermi energy,
$m_n$ is the number of TRI points enclosed by a band $n$ and $|\psi_n(k_{F,n})\rangle$ is the eigenstate on band $n$ at the Fermi surface. In TIs of the Bi$_2$Se$_3$ family
the only TRI point enclosed by the surface bands is the $\Gamma$ point so that $m_n=1$. On the other hand, due to the gap isotropy Eq. (\ref{invariant}) can be
evaluated along any direction in the $k_x-k_y$ plane.

As a paradigmatic example we shall examine the case $N_z=2$. Details 
on the calculations are presented in the SM \cite{supplementary}, where we also discuss the peculiar $N_z=1$ case.
The spectrum for $N_z=2$ consists of four bands with positive energy which, expanded in $\bf{k} \equiv \bf{k}_{\parallel}$,  are given by
\begin{equation}
E_{\alpha,\chi}(k) = \sqrt{E_1^2+2 \alpha F_{\chi }+A_{\chi}^{2}+V^{2}} \text{.}
\label{Bands}
\end{equation}
$\alpha=\pm 1$ is a band index, $E_1^2 = \epsilon_k^2+A^2k^2 + B^2 + C^2$,  $F_{\chi } =\sqrt{\left( BC-\chi A_{2}\left\vert k \right\vert V\right)^{2}+\epsilon_k^{2} (V^2+ B^2) }$, 
$\epsilon_{k}=M_{0}-2B_{1}/a^{2}+B_{2}k^{2}$, with $k=|{\bf k}|$ and we have defined the parameters
as $A=A_1/a$, $B=B_1/a^2$ and $C=A_2/2a$. The bands and their evolution with voltage $V$ are shown in Fig \ref{figure1}(a). We focus on a chemical potential $\mu$ as indicated in Fig. \ref{figure1}(a), intersecting the bands with
$\alpha=-1$. 
A non-trivial value of the $Z_2$ invariant in the present case (i.e. $N=-1$)
implies simply different signs of the projected pairing in the two helicity channels,
\begin{equation}
\left\langle \psi_{\chi} |{\cal T} \hat{\Delta}^{\dag }|\psi_{\chi} \right\rangle =2|D_+|^2\left( \Delta _{+}+\Delta _{-}\pi _{\chi }^{2}\right) 
\left( 1-\beta_{\chi }\Lambda \right).  \label{invar}
\end{equation}
In this expression we have introduced the quantities $D_+$, $\pi_{\chi}=D_-/D_+$ and $\beta_{\chi}=  2 \pi_{\chi}/(\Delta _{+} + \Delta _{-}\pi_{\chi}^2)$, which are defined 
from the  components of the eigenstates of  $\hat{H}^{TB}$ on the bottom surface, i.e. we have $\left\vert \psi _{\chi}\right\rangle =\left( \hat{D}_{\chi },\hat{U}_{\chi }\right) ^{T}$, where $\hat{U}_{\chi}=(U_+,U_-)^T \otimes \hat{\phi}_{\chi}$ and $\hat{D}_{\chi}=(D_+,D_-)^T\otimes \hat{\phi}_{\chi}$, and $\hat{\phi}_{\chi}$ are the eigenstates of the helicity operator, so that $\phi_{\chi}$ measures the relative weight of the two parity sector components on the bottom surface.
We then see that for having a non-trivial value of the $Z_2$ topological  invariant, the necessary (however not sufficient) condition is $\pi_{\chi}$ (or equivalently $\beta_{\chi}$) having different signs for the two helicities. An analytic expression for $\pi_{\chi}$ is given in \cite{supplementary}. 

\begin{figure}[t]
\includegraphics[width=\columnwidth]{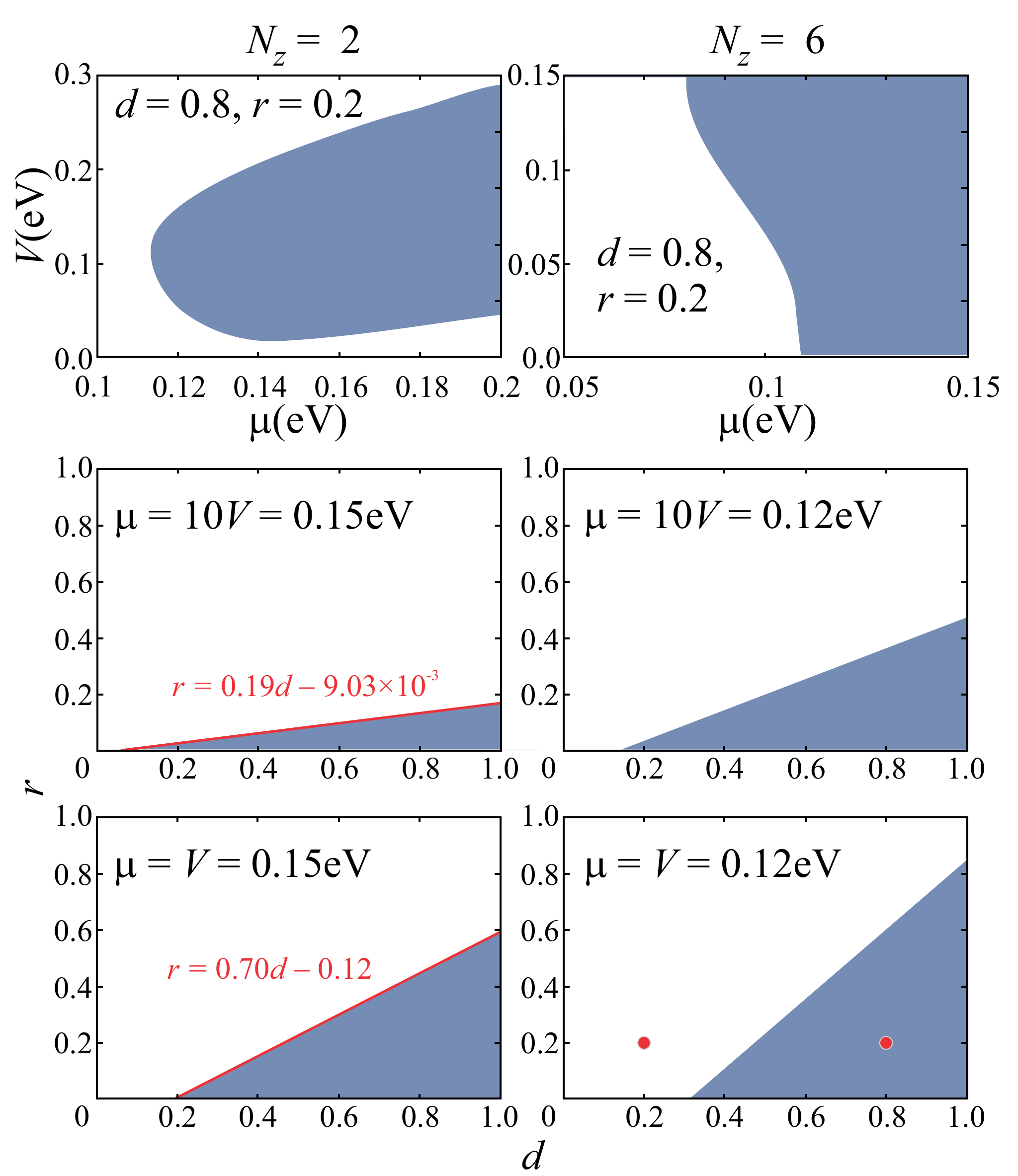}
\caption{Phase diagrams in the $V$, $\mu$ plane at fixed $r=0.2$ and $d=0.8$ (upper panels) and in the
$r=\Delta_+/\Delta_-$, $d=\Lambda/\Delta_-$ plane at fixed 
$\mu=10V$ (middle panels) and at $\mu=V$ (lower panels) 
for the cases $N_z=2$  and $N_z=6$.
The dark (white) color indicates the topological (trivial) regions. As can be observed,
larger values of $V$ help to stabilize the topological phase for in a broader parameter region. The red lines in the $d-r$ diagrams for the $N_z=2$ case are the analytical prediction for the phase boundary as described in the SM \cite{supplementary}.}
\label{figure2}
\end{figure}

As can be observed in Fig. \ref{figure1}(a), the $\pi_{\chi}$ parameter evolves
differently along the lowest bands with opposite helicities, which are split
due to the action of the electric field. While it remains
negative for the $\chi=-1$ band for all values of the chemical potential within the TI gap, in the $\chi=+1$ band it evolves from negative to positive above a certain critical value of the momentum.  As a consequence, for a chemical potential within this energy range and depending on the bias potential $V$, the  effective pairing of Eq. (\ref{invar}) may have different signs on the two helicity bands leading to a non-trivial value of the $Z_2$ invariant, provided that, in addition, 
$\beta_+ > 1/\Lambda$. (see SM \cite{supplementary} for further  details).
In the following we study the occurence of the TS phase as a function of the parameters $r=\Delta_+/\Delta_-$ and $d=\Lambda/\Delta_-$ which determine the relative size of the intra and interparity pairing. We take $\Delta=\Delta_-$ as the reference energy.

In the Fig. \ref{figure2} we show the phase diagrams in the $(\mu,V)$ and in the 
$(d, r)$ planes for the $N_z=2$ and $N_z=6$ cases. 
As can be observed in the upper panels, the topological phase appears for $\mu$ above a certain value which decreases for increasing $N_z$, corresponding to the closing of the hibridization gap between the surface states. 
On the other hand $\mu$ should not exceed
a value $\sim 0.25 eV$ where higher bands start to be populated. 
In addition we observe that a finite $V$ is needed in order to extend the stability
of the topological phase in the $(d, r)$ plane. For small $V$ values (middle panels in Fig. \ref{figure2}) the stability is restricted to the regions $d \rightarrow 1$ and
$r \rightarrow 0$ but these regions grow when $V \sim \mu$, gradually reaching
the optimal case where the TS phase appears for $\Lambda > \sqrt{\Delta_+\Delta_-}$.

\begin{figure}[t]
\includegraphics[width=\columnwidth]{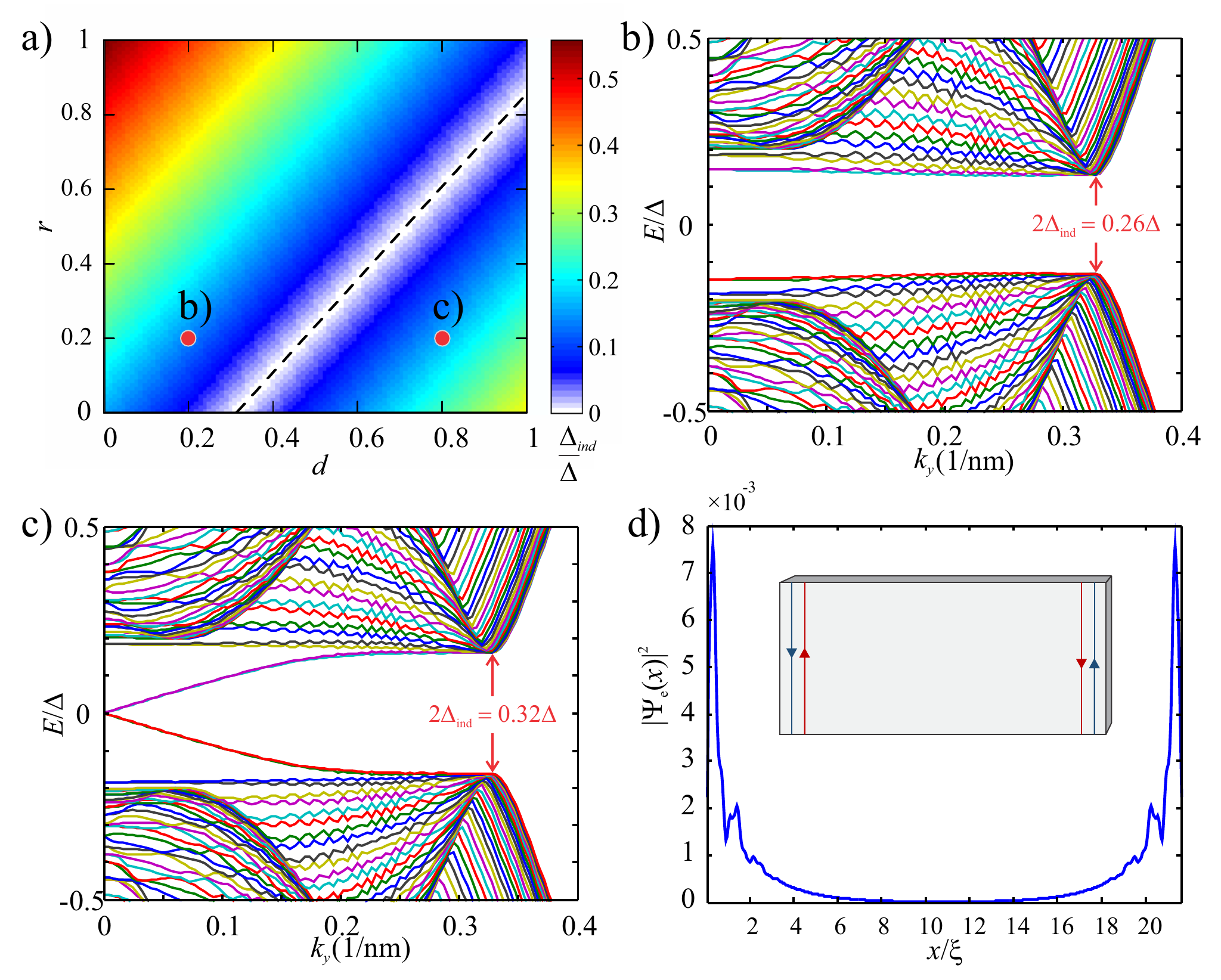}
\caption{a) Induced gap $\Delta_{ind}$ in the $(d, r)$ plane for the same parameters
as in the lower right panel of Fig. \ref{figure2}.
b-c) BdG spectrum for a slab of finite width $W \sim 20\xi$, where $\xi \sim A_2/\Delta$ is the coherence length in the TI film
for the cases indicated by the red dots in panel a). As can be observed, subgap states reaching zero energy for $k_y \rightarrow 0$ appear in the topological case. 
b) Electron probability amplitude for these zero energy states.}  
\label{figure3}
\end{figure}

Another important aspect of the proximity effect in the TI film is the size of the
induced gap parameter $\Delta_{ind}$, which is determined by the smaller
value of $|\left\langle \psi_{\chi} |{\cal T} \hat{\Delta}^{\dag }|\psi_{\chi} \right\rangle|$ at the Fermi surface. As shown in
Fig. \ref{figure3}(a) this quantity drops to zero at the boundary between the 
trivial and the TS region at the $(d,r)$ plane, as expected for a topological transition, and increases when departing from this boundary. The size of $\Delta_{ind}$ can be more clearly appreciated in Figs. \ref{figure3}(b,c) where we show the BdG spectrum for a $N_z=6$ film of finite width in the $x$ direction. The two cases correspond to parameters within 
the trivial and the topological regions, as indicated by the two dots in
the lower right panel of Fig. \ref{figure2} and in Fig. \ref{figure3}(a). In the former case the spectrum exhibits a pair subgap states, droping to zero energy for $k_y \rightarrow 0$. The corresponding wavefunction, exhibiting localization at the edges of the film, 
is plotted in Fig. \ref{figure3}(d). Notice that the localization length is of the order of $5\xi$, which coincides with an effective coherence length $\xi_{eff} \sim A_2/\Delta_{ind}$. As expected for a TS-DIII phase, these states
correspond to Kramers pairs of Majorana modes. 

\begin{figure}[t!]
\centering
\includegraphics[width=0.9\columnwidth]{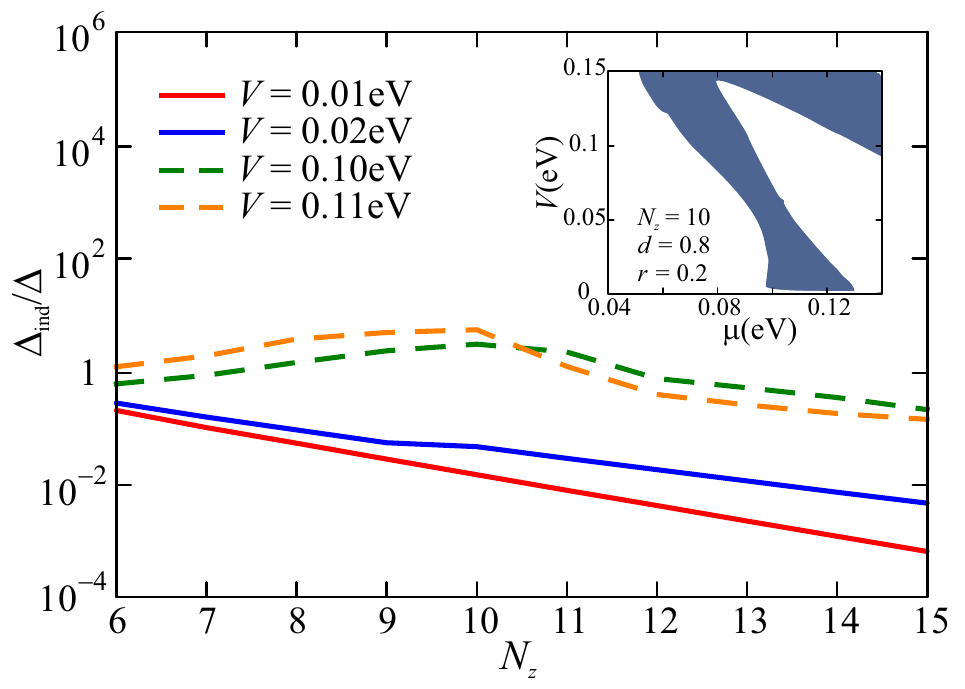}
\caption{Induced gap as a function of the TI layer thickness ($N_z)$ for 
fixed $\mu=0.04 eV$ and different values of $V$. Full and dashed lines indicate trivial and topological phases respectively. The
relative weight of the inter and intra parity pairing at the bottom layer was fixed 
to $d=0.8$ and $r=0.2$. The inset shows the phase diagram in the $\mu$-$V$
plane for $N_z=10$.}
\label{figure4}
\end{figure}

The above results correspond to very thin TI layers with $N_z \le 6$. When 
$N_z$ is further increased the DIII-TS phase can still be reached for 
certain parameters values, but the topological region shrinks and
the phase diagram starts to exhibit disconnected regions, as shown in the inset
of Fig. \ref{figure4} for $N_z=10$. The behavior of the induced gap with $N_z$ depends on the $V$ value. While for small $V$ values, i.e., outside the topological region, it decreases exponentially with $N_z$; for larger $V$ it exhibits a non-monotonic behavior, first increasing and eventually decreasing for $N_z \ge 10$, as can be observed in the main
frame of Fig. \ref{figure4}. This behavior is associated to the fact that higher bands start to cross the chemical potential. 

A word of caution is in order regarding the reliability of the precise quantitative predictions of our model, which can nevertheless be trusted at a qualitative level.
It is also important to remark that when the film is contacted to a superconductor through both surfaces, thus recovering the inversion symmetry, the topological phase disappears \cite{supplementary}. 

\section{Conclusions} 

We have shown that a time reversal invariant TS phase can be induced on a TI film of the Bi$_2$Se$_3$ family
proximitized by a conventional superconductor. In contrast to previous proposals our mechanism does not rely on tuning the phase difference in a S/TI/S junction but arises from the induced interparity pairing which naturally
occurs at the S/TI interface. The mechanism requires breaking the spatial inversion symmetry and a certain degree of hybridization between the TI surface states. 
Application of an electric field perpendicular to the layers helps to stabilize the TS phase for thicker fims in a broader parameter space. Notice that such fields appear spontaneously at the interface between a TI film and its substrate due to charge accumulation \cite{Bi2Se3-films}.

As a final remark let us mention that proximitized Bi$_2$Se$_3$ films have been analyzed in several
experiments, either through Josephson effect \cite{goldhaber,brinkman} or by tunnel spectroscopy \cite{yang,finck}. 
We hope that our work could motivate further experimental studies on this type of devices. 

{\it Acknowledgments:}  We thank L. Brey and J. Schmalian for useful comments on the manuscript. 
This work has been supported by Spanish MINECO through Grants No.~FIS2014-55486-P, FIS2017-84860-R and through the ``Mar\'{\i}a de Maeztu'' Programme for Units of Excellence in R\&D (MDM-2014-0377).
LA thanks support from CONICET and SECyT from Argentina, as well as the Alexander von Humboldt foundation from Germany. WJH and OEC acknowledge funding from COLCIENCIAS, project No. 110165843163 and doctorate Scholarship 617.

\section{Supplemental material}
	
	\title{Supplemental material: Proximity induced time-reversal topological
		superconductivity in Bi$_2$Se$_3$ films without phase tuning.}
	\author{Oscar E. Casas}
	\author{Liliana Arrachea}
	\author{William J. Herrera}
	\author{Alfredo Levy Yeyati}
	\maketitle
	
	\affiliation{Departamento de F\'{\i}sica, Universidad Nacional de Colombia, Bogot\'a,
		Colombia} 
	\affiliation{Departamento de F{\'i}sica Te{\'o}rica de la Materia
		Condensada C-V, Condensed Matter Physics Center (IFIMAC) and Instituto
		Nicol\'as  Cabrera,  Universidad Aut{\'o}noma de Madrid, E-28049 Madrid,
		Spain}
	
	\affiliation{International Center for Advanced Studies, Escuela de Ciencia y
		Tecnolog\'{\i}a, Universidad Nacional de San Mart\'{\i}n-UNSAM, Av 25 de
		Mayo y Francia, 1650 Buenos Aires, Argentina}
	
	\affiliation{Departamento de F\'{\i}sica, Universidad Nacional de Colombia, Bogot\'a,
		Colombia}
	
	\affiliation{Departamento de F{\'i}sica Te{\'o}rica de la Materia Condensada
		C-V, Condensed Matter Physics Center (IFIMAC) and Instituto Nicol\'as 
		Cabrera,  Universidad Aut{\'o}noma de Madrid, E-28049 Madrid, Spain}
	
	\subsection{Unitary transformation on the $\mathbf{{k}{\cdot}{p}}$ model of
		Ref. [1]}
	
	According to Ref. \cite{kp-Zhang} the low-energy and long-wavelength
	electronic properties of Bi$_2$Se$_3$ can be described by a Hamiltonian
	written in a basis of four states which are combinations of $p_z$ orbitals
	of Bi and Se with even and odd parities given by
	\begin{equation}
	\tilde{H}^{3D}= \mathcal{M}(\mathbf{k})\tau_z\otimes \sigma_0 + A_1 k_z \tau_x
	\otimes \sigma_z + A_2 \left(k_x \sigma_x + k_y \sigma_y\right) \otimes
	\tau_x \;,  \label{H-Zhang}
	\end{equation}
	where $\mathcal{M}(\mathbf{k}) = M_0 - B_1 k_z^2 - B_2 (k_x^2+k_y^2)$ and 
	$\tau_{\alpha}$ and $\sigma_{\alpha}$ are Pauli matrices
	operating in different pseudospin spaces. While matrices $\tau_{\alpha}$
	refer to the P1$_z^+$ and P2$_z^-$ orbitals which are mainly located on the
	Bi and Se sublattices respectively \cite{kp-Zhang}, the $\sigma_{\alpha}$
	matrices do not correspond exactly to the \textit{real} spin but are related
	to it by \cite{Silvestrov}
	\begin{equation}
	s_x = \tau_z \otimes \sigma_x, \; s_y = \tau_z \otimes \sigma_y, \; s_z =
	\tau_0 \otimes \sigma_z \;.
	\end{equation}
	For analyzing proximity induced superconductivity it is convenient to write
	the system Hamiltonian in a basis where the $\sigma_{\alpha}$ matrices
	correspond to the real spin. For this purpose one can perform the following
	unitary transformation \cite{Silvestrov}
	\begin{equation}
	U = \frac{\tau_0 + \tau_z}{2} + i \left(\mathbf{\sigma \cdot n} \right) 
	\frac{\tau_0-\tau_z}{2} \;,
	\end{equation}
	where $\mathbf{n}$ is a unitary vector along the $c$ axis. Applying this
	transformation to Hamiltonian (\ref{H-Zhang}) for $\mathbf{n} = +\hat{%
		\mathbf{z}}$ one obtains the model given in Eq. (1) of the main text.
	
	\subsection{Origin of intra and inter parity pairing}
	
	The description of the superconducting proximity effect used in this work can be considered as the low-energy theory arising from integrating out the degrees of freedom of the proximitized superconductor in the Green's function formalism. After this integration the proximity effect is encoded in a self-energy \cite{stanescu}
	\begin{equation}
	\hat{\Sigma}_{\tau,\tau'}(\omega) = \hat{t}_{\tau} \hat{g}(\omega) \hat{t}_{\tau'} \;
	\end{equation}
	where $\hat{t}_{\tau}= t_{\tau} \eta_z \sigma_0$, where $t_{\tau}$ are the hopping amplitudes from the SC to the $\tau$ parity orbitals on the first layer of the TI, $\eta_j$ denote the Pauli matrices in particle-hole space, and $\hat{g}(\omega)$ is the superconductor Green function, which for the BCS case is given by
	\begin{equation}
	\hat{g}(\omega) = \pi\rho_F \frac{-\omega + \Delta_0 \eta_x}{\sqrt{\Delta_0^2 - \omega^2}}\sigma_0 \;,
	\end{equation}
	where $\Delta_0$ is the pairing potential of the parent superconductor and $\rho_F$ denotes its Fermi level density of states. 
	The parameters $\Delta_{\pm}$ and $\Lambda$ in the main text thus arise from the elements $\Sigma^{e,h}_{\tau,\tau'}$ in the limit $\omega \rightarrow 0$.  
	
	\subsection{Limit $N_z=1$}
	
	The aim of this section is to show that the model we are considering
	contains the ingredients for topological superconductivity even in the
	extreme limit of a single monolayer in proximity to an ordinary
	superconductor. 
	
	We consider the  Hamiltonian  of Eq. (2) of the main text with $N_z=1$ 
	\begin{equation}
	\hat{\mathcal{H}}_{1} =\epsilon ({\bf k}_{\parallel})\tau _{z}
	+ a_{2}\tau _{x}\otimes \left[ \sin (k_{y}a)\sigma _{x}-\sin (k_{x}a)\sigma
	_{y}\right] ,   \label{hz1}
	\end{equation}%
	where we have defined $a_{2}=A_{2}/a$. We introduce the Nambu spinor $\Psi
	^{\dagger }(\mathbf{k}_{\parallel })=\left( \psi ^{\dagger }(\mathbf{k}%
	_{\parallel }),-i\sigma _{y}\psi (-\mathbf{k}_{\parallel })\right) $, where $%
	\psi (\mathbf{k}_{\parallel })$ is a spinor of the basis of $\hat{\mathcal{H}}_{1}$. 
	
	The matrix for the Bogoliubov de Gennes Hamiltonian of the monolayer of
	the Hamiltonian of Eq. (3) of the main text with induced superconductivity
	reads  
	\begin{eqnarray}\label{bdg1}
	\hat{\mathcal{H}}_{1}^{BdG} &=& \left( \hat{\mathcal{H}}_{1} - \mu\right)\otimes \eta _{z} +  \\
	& + &\left[\frac{ \Delta_{+} }{2} \left(\tau_0+ \tau_z \right)+\frac{ \Delta_{-} }{2} \left(\tau_0- \tau_z \right) + \Lambda\tau _{x}\right]\otimes  \eta _{x} . \nonumber
	\end{eqnarray}%
	We can now introduce a rotation with respect to the $y$ axis in the
	parity space such that $\tau _{x}\rightarrow -\tilde{\tau}_{z}$ and $\tau
	_{z}\rightarrow \tilde{\tau}_{x}$ and focus on the case with $\Delta_+=\Delta_-=0$. The resulting Hamiltonian reads 
	\begin{eqnarray}
	\tilde{\mathcal{H}}_{1}^{BdG} &=&\eta _{z}\otimes \left( \epsilon
	({\bf k}_{\parallel})\tilde{\tau}_{x}-\mu \tau _{0}\right) -\Lambda \left( \eta
	_{x}\otimes \tilde{\tau}_{z}\right)   \notag \\
	&&-a_{2}\tilde{\tau}_{z}\otimes \left[ \sin (k_{y}a)\sigma _{x}-\sin
	(k_{x}a)\sigma _{y}\right] \otimes \eta_z .  \label{HZ1p}
	\end{eqnarray}%
	
	We  see that along the directions $(k_{x},0)$ and $(k_{y},0)$ this
	Hamiltonian has an identical structure as the 1D model  ladder model introduced by Keselman et al in
	Ref. \onlinecite{kesel}, upon identifying the two parity projections $e,o$
	(even, odd) with the two legs of the ladder Hamiltonian of that paper.
	Importantly, the Hamiltonian of Eq. (\ref{HZ1p}) has different signs of the
	pairing and of the spin orbit interaction in the two different parity
	channels of the rotated basis. Therefore, we know from Ref. \onlinecite{kesel} that it hosts a topological phase.

	\subsection{Model and symmetries for $N_{z}=2$}
	
	Here we consider the case $N_{z}=2$, for which we can obtain some analytical
	results, specially for the $\mathbb{Z}_{2}$ topological invariant. The two layers are labeled with the indices $U$ (up) and $D$
	(down). The matrix for the model Hamiltonian can be written as follows
	
	\begin{eqnarray}
	\hat{\mathcal{H}}_{2} &=&\epsilon (\mathbf{k}_{\parallel })f_{0}\otimes \tau
	_{z}\otimes \sigma _{0}+A_{2}f_{0}\otimes \tau _{x}\otimes k_{\parallel }%
	\hat{h}  \label{HZ2} \\
	&&+Bf_{x}\otimes \tau _{z}\otimes \sigma _{0}+Cf_{y}\otimes \tau _{y}\otimes
	\sigma _{0}  \notag \\
	&&-Vf_{z}\otimes \tau _{0}\otimes \sigma _{0}\text{,}  \notag
	\end{eqnarray}%
	where $B=B_{1}/a^{2}$, $C=A_{1}/2a$, $f_{i}$ are Pauli matrices in the
	surface space, and $\epsilon =M_{0}-2B_{1}/a^{2}-B_{2}\left\vert \mathbf{k}%
	\right\vert ^{2}$, while $V$ is the
	scalar potential, representing a potential difference of $2V$ between the
	opposite surfaces of the film.
	
	We introduce the discretized version of the helicity operator $\Xi=f_0 \otimes \tau _{0}\otimes \hat{h}$. The Hamiltonian (\ref{HZ2}) commutes with this operator, $\left[ \Xi ,\hat{\mathcal{H}}_{2}\right] =0$. Thus, the states can be
	labeled with $\chi =\pm $, which correspond to the two eigenstates of the
	helicity operator $\hat{h}$, $\hat{\phi}_{\chi }(\mathbf{k}_{\parallel
	})=\left( i\chi ,e^{i\theta }\right) ^{T}/\sqrt{2}$, with $e^{i\theta
	}=\left( k_{x}+ik_{y}\right) /k_{\parallel }$.
	
	The Hamiltonian (\ref{HZ2}) is time-reversal invariant, i.e. $\mathcal{T\hat{%
			H}}\left( \mathbf{k}_{\parallel }\right) \mathcal{T}^{-1}=\mathcal{\hat{H}}%
	\left( -\mathbf{k}_{\parallel }\right)$. 
	For $V=0$ it also has inversion symmetry $\mathcal{P\hat{H}}\left( \mathbf{k}%
	_{\parallel }\right) \mathcal{P}=\mathcal{\hat{H}}\left( -\mathbf{k}%
	_{\parallel }\right) $ with $\mathcal{P}=f_{x}\otimes \tau _{z}\otimes
	\sigma _{0}$. Therefore, the eigenstates with eigenenergies $E_{n,\mathbf{k}%
		_{\parallel },\chi }$ satisfy the following properties
	
	\begin{eqnarray}
	\mathcal{T}\psi _{n,\mathbf{k}_{\parallel },\chi } &=&\psi _{n^{\prime }-%
		\mathbf{k}_{\parallel },\chi }\text{,} \\
	E_{n,\mathbf{k}_{\parallel },\chi } &=&E_{n^{\prime },-\mathbf{k}_{\parallel
		},\chi }\text{,}
	\end{eqnarray}%
	and for $V=0$%
	\begin{eqnarray}
	\mathcal{P}\psi _{n,\mathbf{k}_{\parallel },\chi } &=&\psi _{n,-\mathbf{k}%
		_{\parallel },-\chi }\text{,} \\
	E_{n,\mathbf{k}_{\parallel },\chi } &=&E_{n,-\mathbf{k}_{\parallel },-\chi }%
	\text{.}
	\end{eqnarray}
	
	We also consider the Bogoliubov de Gennes Hamiltonian for the bilayer system
	in proximity to superconductors in the most general case, where the induced
	pairing potential has different amplitudes in the two layers, $\hat{\Delta}%
	^{U,D},\;\hat{\Lambda}^{U,D}$. The corresponding matrix reads
	
	\begin{equation}
	\hat{\mathcal{H}}_{2}^{BdG}=\mathcal{H}_{2}\otimes \eta _{z}+\mathcal{H}%
	_{\Delta },  \label{bdg}
	\end{equation}%
	with 
	\begin{equation*}
	\mathcal{H}_{\Delta }=\eta _{x}\otimes \left( \overline{\Delta }+\tilde{%
		\Delta}f_{z}+\left[ \overline{\Lambda }+\tilde{\Lambda}f_{z}\right] \otimes
	\tau _{x}\right) ,
	\end{equation*}%
	where $\overline{\Delta }=\left( \hat{\Delta}_{U}+\hat{\Delta}_{D}\right)
	/2,\;\overline{\Lambda }=\left( \hat{\Lambda}_{U}+\hat{\Lambda}_{D}\right) /2
	$ and $\tilde{\Delta}=\left( \hat{\Delta}_{U}-\hat{\Delta}_{D}\right) /2,\;%
	\tilde{\Lambda}=\left( \hat{\Lambda}_{U}-\hat{\Lambda}_{D}\right) /2$, where
	the pairings $\hat{\Delta}^{U,D}$ in the parity basis are given by $\hat{%
		\Delta}^{U,D}=\Delta _{+}^{U,D}(\tau _{0}+\tau _{z})/2+\Delta
	_{-}^{U,D}(\tau _{0}-\tau _{z})/2$.
	
	In addition to time-reversal, this model Hamiltonian exhibits charge
	conjugation, $\mathcal{C\mathcal{\hat{H}}}_{2}\left( \mathbf{k}_{\parallel
	}\right) \mathcal{C}^{-1}=-\mathcal{\mathcal{\hat{H}}}_{2}\left( -\mathbf{k}%
	_{\parallel }\right) $ with $\mathcal{C}=\eta _{x}\otimes f_{0}\otimes \tau
	_{0}\otimes \sigma _{0}K$, and chiral symmetry $\Pi \mathcal{\hat{H}}%
	_{2}\left( \mathbf{k}_{\parallel }\right) \Pi ^{-1}=-\mathcal{\hat{H}}%
	_{2}\left( \mathbf{k}_{\parallel }\right) $ with $\Pi =\mathcal{CT}$. This
	means that this system belongs to the Altland-Zirnbauer class DIII. For $V=0$ and 
	$\tilde{%
		\Delta}=\tilde{\Lambda}=0$, it has, in addition, inversion symmetry with $%
	\mathcal{P}=\eta _{0}\otimes f_{x}\otimes \tau _{z}\otimes \sigma _{0}$.
	
	\subsection{$\mathbb{Z}_{2}$ topological invariant}
	
	We now consider the Hamiltonian of Eq. (\ref{HZ2}) and focus on the
	direction $\mathbf{k=}\left( 0,k_{y}\right) $. The eigenstates of $\mathcal{H%
	}_{2}$ are expressed as $\left\vert \psi _{\chi }\right\rangle =\left( \hat{D%
	}_{\chi },\hat{U}_{\chi }\right) ^{T}$, with $\hat{U}_{\chi
	}=(U_{+},U_{-})^{T}\otimes \hat{\phi}_{\chi }$ and $\hat{D}_{\chi
	}=(D_{+},D_{-})^{T}\otimes \hat{\phi}_{\chi }$. From the equation $\hat{H}%
	\left\vert \psi _{\chi }\right\rangle =E_{\chi }\left\vert \psi _{\chi
	}\right\rangle $ we get that the components of the spinor $D_{i}$,$U_{i}$
	are real constants satisfying 
	\begin{eqnarray}
	\pi _{\chi ,D} &=&\frac{D_{-}}{D_{+}}  \label{piD} \\
	&=&\frac{\left( CA_{\chi }-X_{+}B\right) Y+\left( CA_{\chi }+Y_{-}B\right)
		\Phi }{\left( BA_{\chi }-X_{-}C\right) Y-\left( BA_{\chi }+Y_{-}C\right)
		\Phi }\text{,}  \notag \\
	\pi _{\chi ,U} &=&\frac{U_{-}}{U_{+}}  \label{piU} \\
	&=&-\frac{\left( BA_{\chi }+Y_{+}C\right) X-\left( BA_{\chi }-X_{+}C\right)
		\Phi }{\left( CA_{\chi }+Y_{-}B\right) X+\left( CA_{\chi }-X_{+}B\right)
		\Phi }\text{,}  \notag
	\end{eqnarray}%
	where $A_{\chi }=A_{2}\chi k $, $Y=A_{\chi }^{2}-Y_{+}Y_{-}$, $X=A_{\chi
	}^{2}-X_{+}X_{-}$, $X_{\pm }=\pm \epsilon \left( k_{y}\right)  -E-V$, $%
	Y_{\pm }=\pm \epsilon \left( k_{y}\right)  -E+V$, $\Phi =C^{2}-B^{2}$.
	The energy bands $E_{\chi }$ are given by the expressions
	\begin{eqnarray}
	E_{\alpha ,\chi } &=&\pm \sqrt{E_1^2+2 \alpha F_{\chi }+A_{\chi}^{2}+V^{2}} \text{,} \nonumber\\
	F_{\chi } &=&\sqrt{\left( BC-\chi A_{2}\left\vert k \right\vert V\right)
		^{2}+\epsilon ^{2} (V^2+ B^2) }\text{.}  \notag
	\end{eqnarray}
	
	Here $\alpha =\pm $ defines a band index and we have introduced the definition
	$E_1^2= \epsilon^2+ B^2 + C^2$. 
	Fig.1 (a) of the main text
	illustrates the behavior of the ratios $\pi _{\chi }$ for different values
	of the potential $V$. Notice that for $V=0$ the bands are degenerate in
	helicity. In fact, for $V=0$ the factor $F_{\chi }$ and then $E_{\chi }$ do
	not depends on $\chi $.

	We now focus on the superconductivity induced only in the $D$ surface, $\hat{%
		\Delta}_{D}=\hat{\Delta},\;\hat{\Delta}_{U}=0$, $\hat{\Lambda}_{D}=\hat{%
		\Lambda},\;\hat{\Lambda}_{U}=0$. This implies $\overline{\Delta }=-\tilde{%
		\Delta}=\Delta /2$, and $\overline{\Lambda }=-\tilde{\Lambda}=\Lambda /2$ in
	Eq.(\ref{bdg}). By calculating the expectation value of $\mathcal{T}\hat{%
		\Delta}^{\dag}$ with the $D$ surface states 
	we obtain Eq. (7) in the main text, i.e.
	\begin{eqnarray}
	\left\langle \psi _{\chi }|\mathcal{T}\hat{\Delta}^{\dag }|\psi _{\chi
	}\right\rangle  &=&2\left\vert D_{+}\right\vert ^{2}\left( \Delta
	_{+}+\Delta _{-}\pi _{\chi }^{2}-2\pi _{\chi }\Lambda \right)   \label{invar}
	\\
	&=&2\left\vert D_{+}\right\vert ^{2}\left( \Delta _{+}+\Delta _{-}\pi _{\chi
	}^{2}\right) \left( 1-\beta _{\chi }\Lambda \right).  \nonumber
	\end{eqnarray}
	Here, we have defined $\pi _{\chi }\equiv\pi _{\chi ,D}\left( E=\mu \right) $ as well as the factor 
	\begin{equation}\label{beta}
	\beta _{\chi }= 2\pi _{\chi }/(\Delta _{+}+\Delta
	_{-}\pi _{\chi }^{2}).
	\end{equation}
	As discussed in the main text, the $\mathbb{Z}_{2}$ invariant of Eq. (5) has a non-trivial value
	when the quantity (\ref{invar})
	has different signs for the $\chi=\pm 1$ bands at the Fermi surface.  
	In turn, a necessary condition is that $\pi_{\chi}$ has opposite signs for the two helicities. If we have $\pi_{\overline{\chi}} <0$ for one band (which warrants
	$\left\langle \psi _{\overline{\chi}}|\mathcal{T}\hat{\Delta}^{\dag }|\psi _{\overline{\chi}}\right\rangle > 0$) the topological phase may exist in a range of parameters satisfying $\pi_{\chi} > 0$ and $\beta_{\chi} > 1/\Lambda$ for the opposite helicity. Notice that
	$\pi_{\pm}$ depends only on the intrinsic parameters, as well as on $\mu$ and $V$, while $\beta_{\pm}$ depends also on the pairing parameters $\Delta_{\pm}$ and $\Lambda$. Hence, the general strategy
	we follow to define the existence of the topological phase is to identify the range of $\mu$ and $V$ for which 
	\begin{equation}\label{bound}
	\beta_{\chi} > 1/\Lambda,\;\;\;\;\;\;\;\; \pi_{\chi} >0,\;\; \pi_{\overline{\chi}}<0.
	\end{equation}
	Notice  that the maximum value
	for $\beta _{\chi }$ occurs for $\pi _{\chi }=\sqrt{\Delta _{+}/\Delta _{-}}$
	and corresponds to $\beta _{\chi }=1/\sqrt{\Delta _{+}\Delta _{-}}$. In such a case, the change of sign in Eq. (\ref{invar}) occurs for $\sqrt{\Delta _{+}\Delta _{-}}<\Lambda$. 
	
	Interestingly, for $V=0$, there are two  limits  where  Eq. (\ref{piD}) simplifies significantly and we can analytically determine the conditions for a topological phase. (i) For $C=0$, the explicit calculation of
	the different coefficients casts $D_+=0$, excluding the possibility of a topological phase. (ii) For $C=B$ we have
	\begin{equation} \label{pic0}
	\pi _{\chi }=-\frac{A_{\chi} - \epsilon + E}{A_{\chi }+ \epsilon + E}%
	,\;\;\;\;\;\;\;\;V=0.
	\end{equation}
	Focusing on $k>0$, we see that $\pi_{+} >0$, while $\pi_{-}<0$ for $E-\varepsilon < A_2 k < E+\varepsilon$. Fixing the chemical chemical potential to satisfy this condition for $k=k_F$ and $E=\mu$, we find the
	range of $\mu$ for the topological phase. Then, given a particular $\mu$ within such range, we can find the conditions to be satisfied for $\Delta_{\pm}, \; \Lambda$ leading to $\beta_{+} > \Lambda$.
	
	\begin{figure}
		\centering	
		\includegraphics[width=\columnwidth]{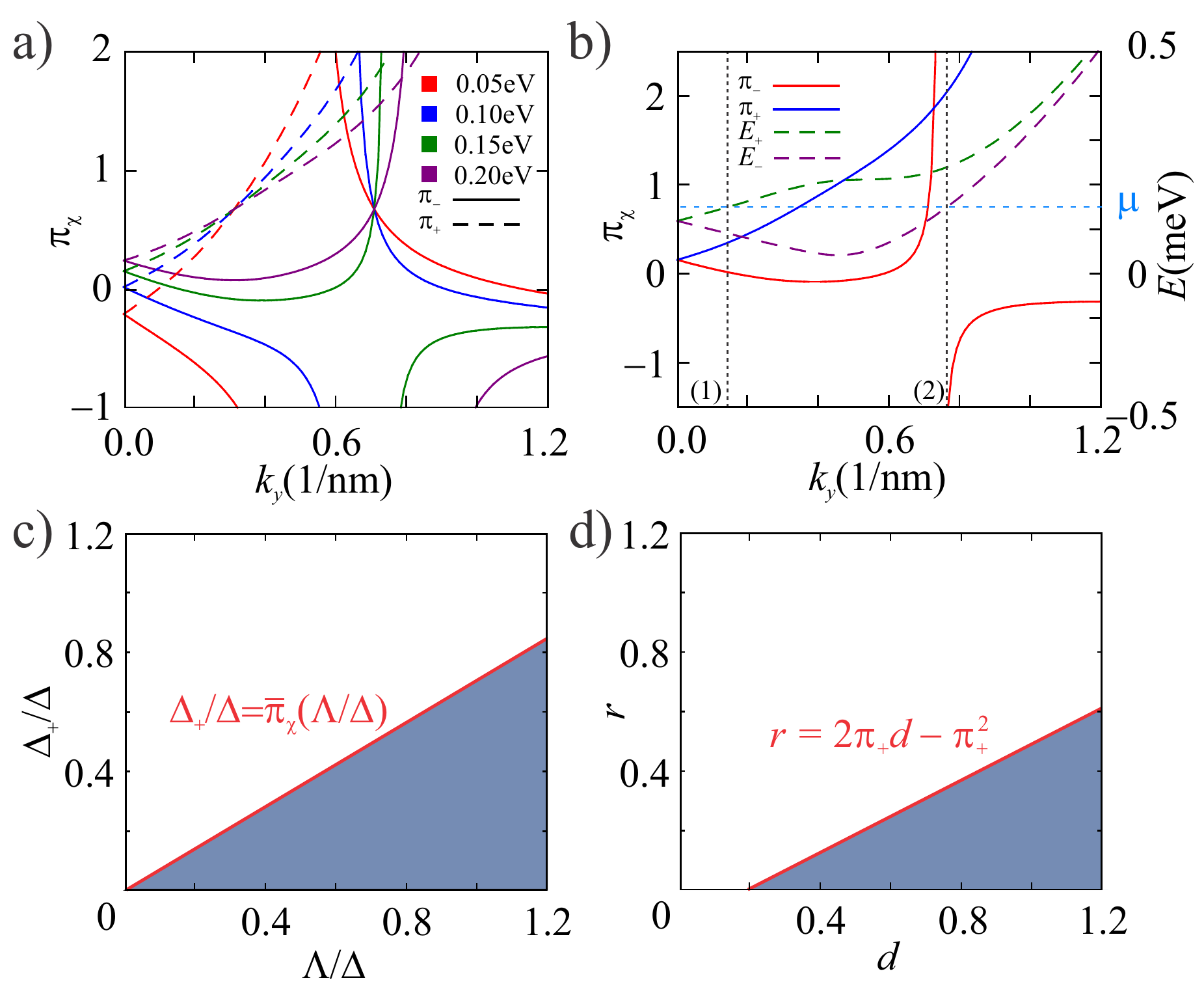}
		\caption{a) Parameters $\protect\pi _{\protect\chi=\pm}$ in the $N_z=2$ case for 
			for several $V$ values as a function of $k_y$. 
			Solid lines correspond to $\pi _{-}$ and
			dashed lines to $\pi _{+}$. b) Parameters $\pi _{\pm }$ (full lines) and energy bands $E_{-1,\pm }\left( k_{y}\right)$ for the case $V=0.15$ eV as a function of $k_{y}$. The vertical dotted lines indicate the Fermi momenta (1) and
			(2) for the case $\mu=V$. 
			c) Phase diagram for the case $\Delta_{+}=\Delta _{-}$. The dark (white) color corresponds to the
			topological (non-topological) phase. 
			d) Same as (c) for the case  $\Delta _{-}=const$.}
		\label{fig:paridad}
	\end{figure}

	In the more general case, it is difficult to draw conclusions on the behavior of 
	$\pi_{\pm}$ on the basis of the analytical expressions, but the procedure is similar than above. Examples are shown in Fig. \ref{fig:paridad} of this Suplemental Material (SM). In panels (a) and (b) Fig. \ref{fig:paridad} the behavior of $\pi_{\pm}$ as a function of $k_y$ is shown for different values of $V$. We see that for most of the cases 
	$\pi_+ >0$ (notice that only in the case with $V=0.05$ eV this is not satisfied close to $k_y=0$). Instead $\pi_-$  shows a discontinuity with a vertical asymptote at $k_y=k_c$, which implies a change of sign at $k_c$. In panel (b) we focus on $V=0.15$ eV and we superimpose the plots of the energy bands. In this way, we can easily identify the portions of the dispersion relation where the conditions 
	$\pi_+>0$ and $\pi_-<0$ are simultaneously satisfied for a certain value of the chemical potential $\mu$.
	
	Panels (c) and (d) in Fig. \ref{fig:paridad} correspond to the phase diagram in the space of the pairing parameters $\Delta_{\pm}$ and $\Lambda$
	for $\mu=V=0.15$ eV. The panel (c) corresponds to $\Delta_+=\Delta_-$ and the boundary separating the topological and non-topological phases is defined from the conditions of Eq. (\ref{bound}), which in this case is given by $\Delta _{+}=2\pi _{\chi }\Lambda/(1+\pi _{\chi }^{2})=\bar{\pi}_{\chi}$. In the panel (b) we illustrate the case with $\Delta _{-}=const$. As in the main text, we define $r=\Delta_+/\Delta_-$ and $d= \Lambda/\Delta_-$. The boundary between the topological and non-topological phases in this case is given by $r=2\pi _{\chi }d-\pi _{\chi }^{2}$. 
	
	For the pairing only in the up-surface we can repeat the previous analysis
	by changing $\pi _{\chi ,D}\rightarrow \pi _{\chi ,U}$, $\Lambda \rightarrow
	-\Lambda $. Finally, in the case of symmetric junction with 
	two identical superconductors contacting the TI film and in the absence of an electric field, the system recovers the inversion symmetry and the topological phase is lost since the contribution to the topological invariant from both surface states has the same sign. While a finite electric field does break this symmetry it does not lead to a robust topological phase, at least for small $N_z$ values, when both surfaces are contacted to a superconductor.

\end{document}